# Hydrodynamic Analysis and Responsivity improvement of a metal/semiconductor/metal plasmonic detector

ELAHE RASTEGAR PASHAKI[1], HASSAN KAATUZIAN [1,*], ABDOLBER MALLAH LIVANI [2]

[1]*Photonics Research Laboratory, Electrical Engineering Department, Amirkabir University of Technology, Tehran 15914, Iran.*
[2]*Electrical Engineering Department, Mazandaran University of Science and Technology, Behshahr, Iran.*
*\* hsnkato@aut.ac.ir*

**Abstract:** Characteristics improvement of photon/plasmon detectors have been the subject of several investigations in the area of plasmonic integrated circuits. Among different suggestions, Silicon-based Metal-Semiconductor-Metal (MSM) waveguides are one of the most popular structures for implementation of high-quality photon/plasmon detectors in infrared wavelengths. In this paper, an integrated Silicon Germanium (SiGe) core MSM plasmon detector is proposed to detect λ=1550 nm with internal photoemission mechanism. Performance characteristics of the new device are simulated with a simplified hydrodynamic model. In a specific bias point (V=3 V and the incident optical power of 0.31 mW), the output current is 404.3 μA (276 μA detection current and 128.3 μA dark current), responsivity is 0.89 A/W and the 3-dB electrical bandwidth is 120 GHz. Simulation results for the proposed Plasmon detector, in comparison with the empirical results of a reported Si-based MSM device, demonstrate considerable responsivity enhancement.

**Keywords:** plasmonics; detector; internal photoemission; SiGe

## 1. Introduction

There are a plethora of researches on finding proper materials for detection of different electromagnetic wavelengths from infrared (IR) to ultraviolet (UV) [1]. However, silicon-based detectors are more desirable because of fabrication technology considerations. The energy band gap ($E_g$) of silicon (Si) is 1.08 eV at room temperature. It means that this material can generate electron-hole pairs (EHP) in visible wavelengths. In order to extend detection wavelengths of Si to IR region, internal photoemission (IPE) mechanism can be replaced instead of EHP. Nevertheless, the detection efficiency of the IPE mechanism is insufficient and plasmons are the key to overcome this limitation [2].

Plasmon detectors have higher absorption rate and are more sensitive to polarization, angle and wavelength of the incident electromagnetic waves. These detectors are commonly made from a metal, which provides the coupling condition of photons to plasmons, and a semiconductor in Schottky junction configuration. Based on the metal architecture in plasmonic detectors, these devices have a wide variety of structures [3-6]. Among those, waveguide plasmonic detectors have particular importance because of CMOS compatibility and integration capability with plasmonic integrated circuit elements [7, 8]. With this in mind, improving characteristics of Si-based waveguide plasmon detectors have been the subject of several investigations [9-11].

In this paper, a Si-SiGe based metal-semiconductor-metal (MSM) waveguide has been proposed as an IR plasmon detector and analyzed with a simplified hydrodynamic model. We have utilized the reported structures of [11, 12] as the basis of our design procedure. In the proposed detector, the lower bandgap of SiGe improves detector's responsivity in comparison with the initial IPE-based Si-core plasmon detector [11]. The process of this paper is as follows: the physical structure of the proposed detector is introduced in section 2. Then, device analysis and simulation method are discussed in sections 3 and 4 respectively. Results of simulations are presented in section 5. Finally, the achievements of this work will be summarized in section 6.

## 2. Device Physical Structure

The physical structure of the proposed plasmon detector is presented in fig. 1. This waveguide has a lightly p-doped Si-SiGe core sandwiched between titanium (Ti) and gold (Au) layers. The photonic to plasmonic mode converter in this detector is a tapered waveguide configuration same as [11] and is not shown in fig. 1. Physical parameters of the detector are listed in table 1. In this structure, carriers flow through the narrower region of the semiconductor core which is considered as an active region (height ≈ 275nm) in fig. 1(a). The maximum height of the SiGe layer is dependent to Ge mole fraction (x) and is determined for x = 5%, 10% and 13% in table 1 [13].

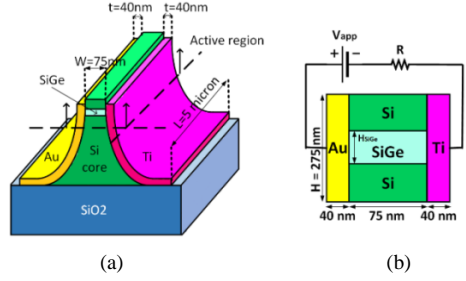

(a)          (b)

Fig. 1. (a) Physical structure and dimensions of the proposed plasmon detector (b) Magnified active region and biasing circuit of the detector.

Table 1. The physical structure of the proposed plasmon detector

|  | Type /dope (cm$^{-3}$) | Width (nm) | Height (nm) | Length (μm) |
|---|---|---|---|---|
| **Si-core** | P-Type / $10^{15}$ | W = 75 | H = 275 | L = 5 |
| **SiGe-core** | P-Type / $10^{15}$ | W = 75 | $H_{SiGe}$ = 200 (x=5%) <br> = 50 (x=10%) <br> = 40 (x=13%) | L = 5 |
| **Metals** | -- | t = 40 |  | L = 5 |

In this structure, the SiGe region has lower $E_g$ in comparison with Si that provides a channel for carriers. Moreover, the lower bandgap of SiGe reduces the height of the Schottky barrier for holes and enhances the dark and detection currents of the detector. In order to compensate the increased dark current, this detector can be imported in a balanced structure, as described in our previous work [14], to isolate output port from the dark current.

### 3. Device Analysis

*3.1 Electromagnetic Analysis*

Mode Analysis of the proposed Au/Si-SiGe/Ti plasmon detector has been done in LUMERICAL simulator for different Ge mole fractions (x). The simulated device structure is shown in fig. 2. This structure consists of a photonic and a tapered waveguide that converts photonic modes to plasmonic ones.

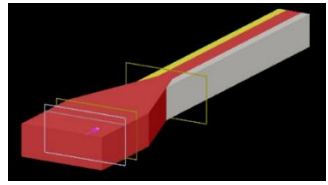

Fig. 2. Simulated structure in LUMERICAL

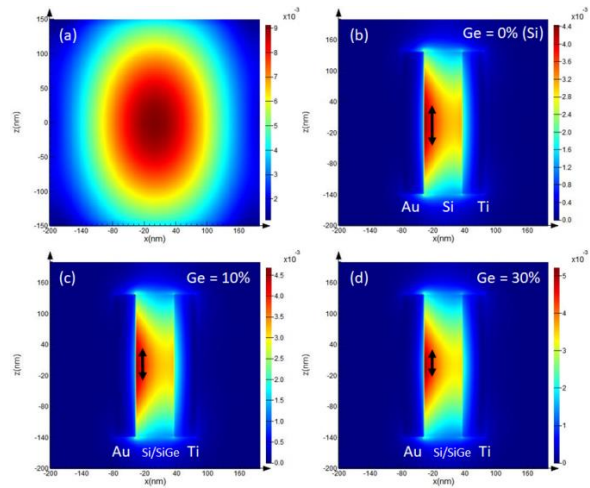

Fig. 3. The propagation mode of a) photonic waveguide and b) plasmonic waveguide with x=0 c) x =10% and d) x=30% SiGe core

The propagation mode of the photonic and plasmonic waveguides with x = 0, 10, 30% are shown in fig. 3. The height of the SiGe layer remains constant (50nm as shown in fig. 1(b)) for all cases to provide a better condition for comparing the x variation effect. As can be seen, by rising the mole fraction, the confinement of electrical field will increase that causes the enhancement of plasmon's loss rate and improving the hot carrier generation in the SiGe region. Moreover, the length of Si-SiGe core MSM can be set much lower than the corresponding structure with Si core that reduces the total area of the proposed detector and can create a positive effect for decreasing the dark current. However, in following simulations we consider the dimensions of Si-SiGe device same as initial Si core structure to have a platform for studying other aspects of SiGe existence in characteristics of the proposed detector.

In another part of LUMERICAL simulations, the coupling coefficient (A) of the tapered waveguide is calculated by dividing the spatial integral of plasmonic wave power into the spatial integral of the photonic wave which is determined as 4% and 3% for Si and SiGe core MSMs respectively. These coupling coefficients will be used in the equation of IPE model in section 3-3 for determining the detection current.

*3.2 Energy band Diagram*

The energy band diagram of the Au-Si-Ti and Au-SiGe-Ti are shown in fig. 4. The Schottky barriers for electrons on each side of this MSM structures can be calculated by $\phi_{Bn}=W-X$, Where W is metal's work function ($W_{Au}$ = 5 eV and $W_{Ti}$ = 4.8 eV) and X is the electron affinity of the semiconductor that considered as 4.2ev for Si and SiGe [11, 15]. Since a Schottky diode operates based on majority carriers, in p-doped core MSM waveguide, all calculations should be done on holes. The Schottky barrier for electrons can be converted to holes by $\phi_{Bp}= E_g-(W-X)$ where $E_{g-si}$ = 1.08 eV and for SiGe can be calculated as [16]:

$$E_g = 1.08 + x(0.945 - 1.08)/0.245 \quad for \quad x \leq 0.245 \tag{1}$$

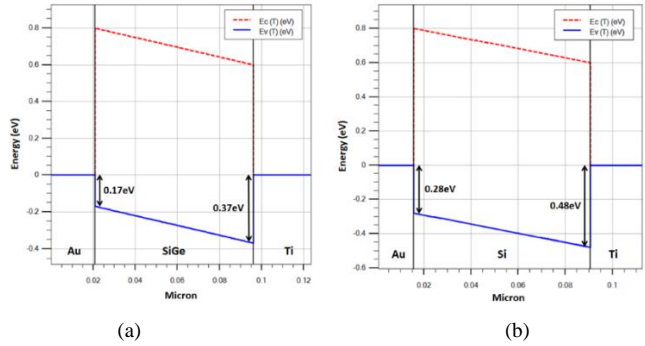

(a)          (b)

Fig. 4. Energy band diagram of (a) Au-SiGe-Ti (b) Au-Si-Ti MSM junctions at room temperature.

The room temperature Schottky barrier height for holes at Au-Si, Au-SiGe, Ti-Si, and Ti-SiGe interfaces are shown in fig.4. To extend these calculations into lower temperatures, we use the equation (2) to determine bandgap energy dependency to temperature [18]:

$$E_g(T) = E_g(0) - \frac{\alpha T^2}{(T+\beta)} = E_g(300K) + \alpha\left(\frac{300^2}{300+\beta} - \frac{T^2}{T+\beta}\right) \tag{2}$$

Where α and β are material dependent coefficients that are 4.73×10-4 eV/K and 636 K for Si [17] and for SiGe are as follows [15]:

$$\alpha = (\alpha_{Si} + x(\alpha_{Ge} - \alpha_{Si}))\times 10^{-4} = (4.73 + x(4.77-4.73))\times 10^{-4} \ eV/K$$
$$\beta = \beta_{Si} + x(\beta_{Ge} - \beta_{Si}) = 636 + x(235-636) \ K \tag{3}$$

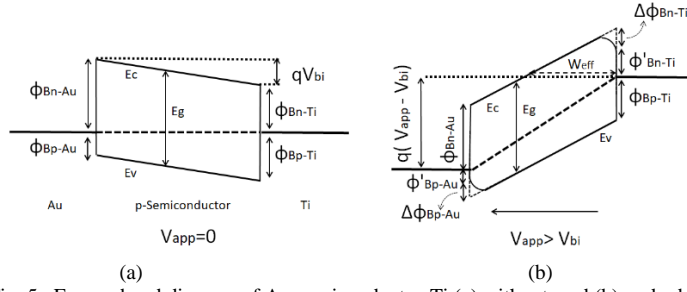

Fig. 5. Energy band diagram of Au-semiconductor-Ti (a) without, and (b) under bias.

Detailed energy band diagram of fig. 4 is sketched in fig. 5(a). According to this diagram, Au contact should be connected to a higher potential than the Ti side to compensate the built-in voltage ($V_{bi}$) arisen from the Schottky barrier difference between two junctions. Under an appropriate biasing condition, energy diagram of fig. 5(a) changes into the fig. 5(b). Under this condition, the Schottky barrier lowering effect changes the height of the barriers. The Schottky effect is the image-force-induced lowering of the potential energy for charge carrier emission and is proportional to applied voltage according to the following equation [17]:

$$\Delta\phi_B = \sqrt{\frac{qV_{app}}{4\pi\varepsilon_s W}} \qquad (4)$$

Here, $V_{app}$ is the applied voltage, q is the elemental charge, $\varepsilon_s$ is semiconductor's permittivity (which is 11.8 for Si and 11.8+4.2x [15] for SiGe) and W is core width in the MSM structure. By applying this effect, in the forward biased junction, the barrier height is slightly larger and for reverse bias, the barrier height becomes slightly smaller than the zero bias condition [17] (e.g. $\Delta\phi_B = 0.0351$ eV at $V_{app} = 2$V).

*3.3 Current Analysis*

Based on the energy diagram of fig. 5(b), the applied voltage causes the flow of holes from gold to titanium contact and creates an electrical current in the same direction. In this section, the current relations of an MSM structure with energy diagram of fig. 5 will be discussed at first. Then, generation of hot holes in the Au side (IPE detection mechanism) will be described which should be imported in current relations for device simulation in detection mode.

3.3.1 MSM current (dark current)

In this paper, we consider simplified hydrodynamic (HD) model [18, 19] to simulate the proposed submicron MSM plasmon detector. This model introduces two independent variables $T_n$ and $T_p$, the temperature of electrons and holes, in addition to lattice temperature ($T_L$) and considers continuity equations for the carrier's temperatures. The HD model equations for holes consist of [19]:

$$\begin{aligned}
div\vec{S}_p &= \frac{1}{q}\vec{J}_p\cdot\vec{E} - W_p - \frac{3K}{2}\frac{\partial}{\partial t}(pT_p) \\
\vec{J}_p &= -qD_p\nabla p - q\mu_p p\nabla\Psi - qpD_p^T\nabla T_p \\
\vec{S}_p &= -K_p\nabla T_p - \left(\frac{K\delta_p}{q}\right)\vec{J}_p T_p
\end{aligned} \qquad (5)$$

Where $S_p$ is hole's energy flux density, $\mu_p$ is hole mobility, E is the electric field, K is Boltzmann constant, $D_p = \mu_p KT_p/q$ is hole's thermal diffusivity, p is hole's concentration and $\psi$ is the electrostatic potential. Other parameters are defined as follows [18]:

$$K_p = qp\mu_p \left(\frac{K}{q}\right)^2 \Delta_p T_p$$

$$\Delta_p = \delta_p \left[\frac{7}{2}\frac{F_0 + 5/2(\eta_p)}{F_0 + 3/2(\eta_p)} - \frac{5}{2}\frac{F_0 + 3/2(\eta_p)}{F_0 + 1/2(\eta_p)}\right]$$

$$\delta_p = \frac{\mu_{2p}}{\mu_p}, \quad \mu_{2p} = \frac{5}{2}\mu_p \frac{F_0 + 3/2(\eta_p)}{F_0 + 1/2(\eta_p)}$$

$$\eta_p = F_{1/2}^{-1}\left(\frac{p}{N_V}\right)$$

$$N_V = \left(\frac{2\pi m_h^* KT_p}{h^2}\right)^{1.5} = \left(\frac{T_p}{300}\right)^{1.5} N_V(300)$$

$$D_p^T = \left(\mu_{2p} - \frac{3}{2}\mu_p\right)\frac{K}{q}$$

(6)

Where F is the Fermi-Dirac integral and $N_V$ is hole's effective density of states. The $W_p$ in eq. (5) is loss rate that includes physical mechanisms by which carriers exchange energy with the surrounding lattice environment and is considered as follows [18]:

$$W_p = \frac{3}{2} p \frac{K(T_p - T_L)}{\tau_h} \tag{7}$$

That $\tau_h$ is holes energy relaxation time in the semiconductor. The holes mobility in previous equations is dependent on temperature and electrical field according to [15]:

$$\mu_p(T) = \mu_p(300)\sqrt{\frac{T_L}{300}} \tag{8}$$

$$\mu_p(E) = \frac{\mu_{p0}}{\sqrt{1 + \left(\frac{\mu_{p0}E}{V_{sat-p}}\right)^2}} \tag{9}$$

Where $V_{sat-p}$ is the hole's saturation velocity and $\mu_p(300)$ and $\mu_{p0}$ are room temperature mobility and low field mobility respectively which are determined based on semiconductor's material and its doping level.

The material parameters in eq. (5) to (9) are for Si and SiGe in simulations of the proposed device. The Si parameters are reported in various resources and some SiGe parameters are considered similar to Si especially in low Ge mole fractions. However, a number of SiGe Parameters have different amounts. For instance, the $N_V$ and holes mobility of SiGe are 0.23 [20] and 1.25 [21] times the Si values respectively for x=10%.

Eventually, solving HD equations needs a boundary condition to determine p and ψ. The boundary condition in the proposed MSM detector is Schottky contact equations. The current of a Schottky junction is divided into thermionic and tunneling parts which are shown for electron carriers in fig. 6.

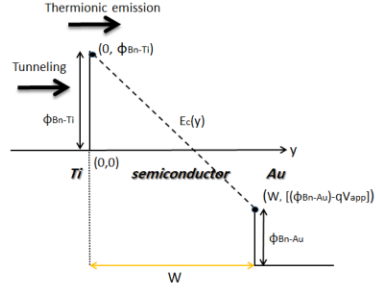

Fig. 6. Detailed energy diagram of Ti-Semiconductor-Au structure

The thermionic component of current in a Schottky junction from metal to semiconductor region is [17]:

$$\vec{J}_{sp} = A_p^* T_L^2 \exp\left(\frac{-q\Phi_{Bp}}{KT_L}\right) \qquad (10)$$

Where $A_P^*$ is effective Richardson's coefficient for holes and $T_L$ is the holes temperature which sets equal to lattice temperature on the contacts. Tunneling component of current in a Schottky junction can be described by [22]: (in the following equations tunneling of electrons are described because of its simpler vision, however, these equations will be adapted to holes at the end of this section.)

$$J_T = \frac{A^* T_L}{K} \int_E^\infty \Gamma(E')\ln\left[\frac{1+fs(E')}{1+fm(E')}\right] dE' \qquad (11)$$

Where $f_s(E)$ and $f_m(E)$ are the Maxwell-Boltzmann distribution functions in the semiconductor and metal, E is the carrier energy and $\Gamma(E)$ is tunneling probability. To obtain the localized tunneling rate ($\Gamma(y)$), eq. (11) is imported in $G_T = (\nabla J_T)/q$ and yields [23]:

$$G_T = \frac{A^* T \vec{E}}{K} \Gamma(y) \ln\left[\frac{1+n/\gamma_n N_c}{1+\exp\left[-(E_c - E_{FM})/KT\right]}\right] \qquad (12)$$

Where E and n are the local electric field and local electron concentration, $N_c$ is the local conduction band density of states, $\gamma_n$ is the local Fermi-Dirac factor, $E_{FM}$ is the Fermi level in the contact and $E_c$ is the local conduction band edge energy which is related to $V_{app}$ according to fig. 6 and can be written as:

$$E_c(y) = \frac{\Phi_{Bn-Au} - \Phi_{Bn-Ti} - qV_{app}}{W} y + \Phi_{Bn-Ti} \qquad (13)$$

The tunneling probability ($\Gamma(y)$) in eq. (12) can be determined by assuming this linear variation of conduction band energy ($E_c$), as follows [23]:

$$\Gamma(y) = \exp\left[\frac{-4\sqrt{2m}\, y}{3\hbar}\left(E_{FM} + \Phi_{Bn-Ti} - E_c(y)\right)^{\frac{1}{2}}\right] \qquad (14)$$

Here, m is the electron effective mass for tunneling and $\hbar$ is reduced Planck's constant. Based on eq. (13), in a specific "y" by increasing the applied voltage, $E_c(y)$ will decrease which causes enhancement of tunneling probability according to eq. (14). Similar expressions of the above equations exist for holes by replacing $E_v(y) = E_c(y)-E_g$, hole's average effective mass, $\phi_{BP}$, etc. instead of corresponding parameters in equations (11) to (14).

3.3.2 IPE Model

After excitation of SPPs in a metal-semiconductor interface, absorption of plasmons can occur in each side of this interface according to the level of photons' energy (E = hυ; h is Planck's constant and υ is optical frequency). For hυ > $E_g$, plasmons absorb in semiconductor side and generate electron-hole pairs (EHP). However, IR photons by free space wavelength of 1550nm (hυ = 0.8 eV) have lower energy than silicon bandgap, so excited plasmons absorb in the metal side and detection occurs based on IPE mechanism. IPE can be described as a 3-step process [2]. 1) Generation of hot carriers by absorption of photons/plasmons in metal side, 2) transmission and scattering of hot carriers toward semiconductor interface, 3) Emission of hot carriers from Schottky barrier and creating detection current. This 3-step process can be described by a semi-classical model [24]. According to this model, internal quantum efficiency is calculated by:

$$\eta_i = \frac{I_p/q}{S_{abs}/h\upsilon} = \frac{1}{2}\left(1 - \sqrt{\frac{\Phi_B}{h\upsilon}}\right)^2 \tag{15}$$

Where $I_p$ is photocurrent and $S_{abs}$ is absorbed optical power, which is converted to incident optical power by $S_{abs}=A\ S_{inc}$ in which A is the merit of photonic to plasmonic mode converter. Therefore, photodetection current in a Schottky interface obtained as follows:

$$I_p = q\frac{AS_{inc}}{2h\upsilon}\left(1 - \sqrt{\frac{\Phi_B}{h\upsilon}}\right)^2 \tag{16}$$

This current should be imported in the dark current of the gold to semiconductor junction (eq. (10)) to determine the total boundary current of the HD model for simulation of the proposed MSM detector in the illumination mode. This process will be described in section 4.

*3.4 Signal and Noise Analysis*

Operating speed limitation of the MSM plasmon detector is relevant to various phenomena, such as hot-carrier lifetime in metals ($\tau_{hc}$), carriers drift time through the semiconductor layer ($\tau_{dr}$) and RC time constant ($\tau_{RC}$) [11], according to the following relation:

$$f_{3dB} = \frac{1}{\sum \tau} \tag{17}$$

The hot carrier lifetime in Au is considered as 30 fs [25]. Due to the narrow thickness (75 nm) of the semiconductor layer, the drift time of carriers is specified by considering saturation velocity which can be calculated for SiGe through [26]:

$$v_{sat-n,p-SiGe}(T,x) = v_{sat-n,p-Si}(T) \times \frac{0.255}{0.255 + x(1-x)} \tag{18}$$

Where $v_{sat-n,p-Si}$ is the saturation velocity of Si at a specific temperature. Accordingly, the hole's saturation velocities are considered as $0.8\times10^7$ cm/s and $0.6\times10^7$ cm/s for Si and SiGe (x=10%) respectively [26]. Finally, the RC time constant of the MSM junction is determined by estimating an equivalent parallel-plate capacitor (C) with 5μm × 275nm metal area across W=75 nm Si core which leads to a capacitance equal to:

$$C_{Si} = \varepsilon_{Si}\varepsilon_0 \frac{L \times H}{W} = 1.9 fF \tag{19}$$

However, in Si-SiGe core MSM, SiGe region is a channel for current due to its lower energy band gap and as it is shown in fig. 7, the effective area will reduce to $L \times H_{SiGe} = 5\mu m \times 50$ nm (for x=10%). Moreover, the relative permittivity of SiGe is [15]:

$$\varepsilon_{Si} = 11.8 + 4.2x \tag{20}$$

Consequently, $C_{SiGe} = 0.36$ fF for x=10%. The load resistance is considered as R = 50 Ω for both MSM devices.

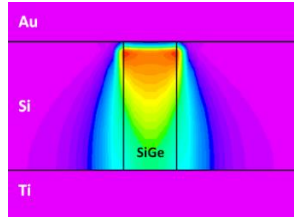

Fig. 7. Current distribution in Au/Si-SiGe/Ti MSM structure.

Given these points, the 3-dB bandwidth of the Si and Si-SiGe core MSM detectors are calculated as 149 GHz and 120 GHz respectively which shows a 19% reduction for the proposed device in comparison with the initial Si-core detector because of lower saturation velocity of holes in SiGe.

Another key characteristic of a detector is the signal to noise ratio (SNR) which can be determined by considering the shot, thermal and dark current noise sources as follows [27]:

$$\frac{S}{N} = \frac{I_{det}^2}{2qB(I_{det} + I_{dark}) + \frac{4KTB}{R} + I_{dark}^2} \tag{21}$$

Where B is the modulation frequency of the input signal. This characteristic will be calculated in section 5.

## 4. Simulation Method

Dark current simulations of the proposed MSM plasmon detector have been done by solving the equations (5) to (14) of section 3. In the detection mode, we use an indirect method for importing detection current (eq. (16)) in Schottky boundary condition of HD calculations. As it was mentioned before, the holes temperatures in contacts are set equal to the lattice temperature and then HD model considers continuity equations for the carrier temperatures. In the detection mode, the average energy of carriers increases by absorbing plasmons energy and generation of hot carriers. Another way to enhance carriers energy (and as a result, improving thermionic emission) is increasing the operational temperature. With this in mind, we use an effective temperature concept to create the detection condition in the Schottky current equation (eq. (10)) and generate hot carriers by increasing the lattice effective temperature ($T_{eff}$) (boundary condition for temperature).

For determining the effective temperatures, the detection currents are calculated for the different incident optical powers (eq. (16)). Then, eq. (10) sets equal to these currents and $T_{eff}$ is calculated for each optical power. The calculated effective temperatures for a number of incident optical powers are listed in table 2. These temperatures have earned a maximum ±5K variations during simulations to create a perfect linear behavior in the responsivity curves but remained constant for each device at different voltages. These calculations have been done on Au-Si junction with $\phi_{Bp}$=0.28eV, $A_P^*$ =30 A/cm$^2$/K$^2$ and λ=1550 nm. These effective temperatures remain constant for simulation of Au/Si-SiGe/Ti structure in each incident optical power.

Table 2. calculated effective lattice temperatures for different incident optical powers

| $S_{inc}$ | 0 (dark mode) | 100 µW | 150 µW | 200 µW | 250 µW | 310 µW |
|---|---|---|---|---|---|---|
| $T_{eff}$ | 250 K | 277 K | 290 K | 299 K | 306 K | 312 K |

In order to verify the validity of the mentioned method, the fabricated structure reported in [11] is simulated and the results are compared with the reported empirical curves in fig. 8. As can be seen, there is a proper agreement between simulated and reported results. Similarities between the structure of the proposed device of this work and reported device of [11], allow us to apply the discussed theories for simulation of the proposed device in the same biasing condition.

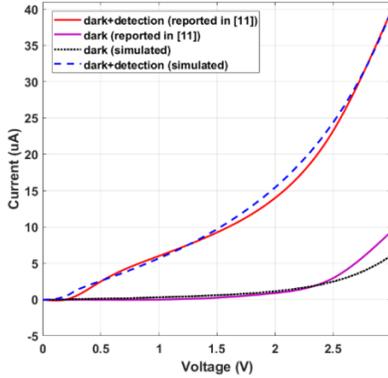

Fig. 8. Comparison of Simulation results with empirical curves [11]. Both experiment and simulation results belong to Au-Si-Ti detector

## 5. Results

The current-voltage characteristic of the proposed MSM plasmon detector is presented in fig. 9. The operational condition of this I-V curve is similar to fig. 8 and as can be seen, the dark and detection currents are 21 and 10 times more than the corresponding values of the initial detector respectively. This dark current enhancement is the cost of increasing the detection current of the plasmon detector.

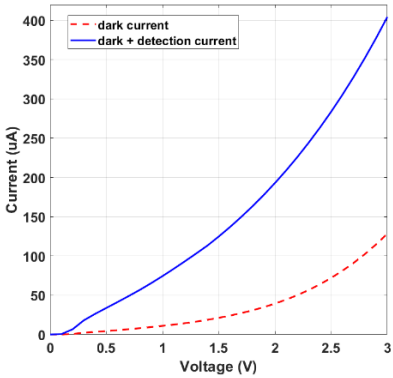

Fig. 9. I-V characteristic of the proposed MSM plasmon detector

The responsivity (R) of a detector is defined as the slope of output current versus optical input power characteristic, which is plotted at three different bias voltages for both Si core and Si-SiGe core devices in fig. 10 and 11 respectively. The responsivity of the proposed detector has ×10.3, ×10 and ×8.5 growth at V = 1, 2, 3 V respectively.

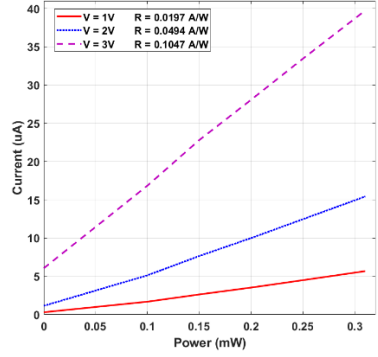

Fig. 10. Output current (detection + dark) as a function of optical input power in Si-core MSM detector for V = 1, 2, 3 V

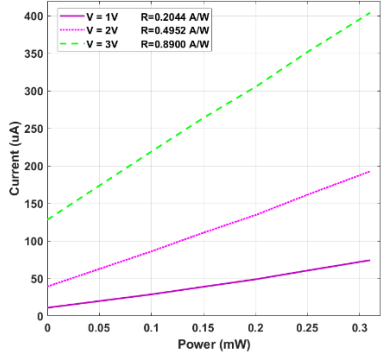

Fig. 11. Output current (detection + dark) as a function of optical input power in Si-SiGe core MSM detector for V = 1, 2, 3 V

The effect of Ge mole fraction variations in the output currents of the proposed plasmon detector is plotted in fig. 12 for x = 5%, 10% and 13% with maximum allowable height for SiGe layer ($H_{SiGe}$ = 200, 50, 40 nm respectively [13]). As can be seen, the dark and detection currents increase very fast by raising the Ge mole fraction.

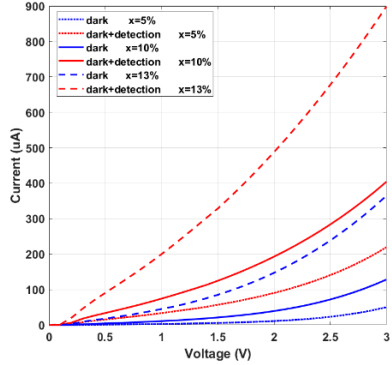

Fig. 12. I-V characteristic of the proposed MSM plasmon detector with x = 5%, 10% and 13%

Finally, in order to evaluate the performance of the proposed device, different parameters of both structures are summarized in table 3 for x = 10%, $S_{inc}$ = 0.31 mW and V = 3V. In Si-SiGe core MSM detector, the dark current noise has the dominant role in comparison with other noise sources and reduces the SNR of the proposed detector. However, by importing these MSM detectors in a balanced structure [14] and eliminating the dark current from the output, the SNR of the proposed detector will increase.

**Table 3. Comparison of simulated parameters for Si-SiGe core and Si core MSM plasmon detectors**
**(@V=3V and $S_{inc}$=0.31 mW)**

|  | Si-SiGe core proposed Detector | Si core initial detector |
|---|---|---|
| **Dark current** | 128.3 μA | 6.07 μA |
| **Output current in detection mode** | 404.3 μA (dark + detection) | 39.74 μA (dark + detection) |
| **Responsivity** | 0.89 A/W | 0.10 A/W |
| **Elec. BW** | 120 GHz | 149 GHz |
| **SNR** | 6.64 dB (@120 GHz) | 11.51 dB (@149 GHz) |
| **SNR In balanced structure** | 38.24 dB | 18.64 dB |

## 6. Conclusion

In this paper, an IPE based Si-SiGe core MSM plasmon detector is proposed and theoretically analyzed. The proposed device has higher detection current and responsivity. However, these advantages are in a trade-off with dark current noise which can be compensated by a balanced structure and electrical modulation bandwidth. Performance of the new device is theoretically investigated with a simplified hydrodynamic model. In a specific bias point (V=3V and $S_{inc}$=0.31 mW), the bandwidth is 120 GHz and SNR is 6.64 dB that have 19% and 42% reduction respectively. However, the output current is 404.3 μA and responsivity is about 0.89 A/W which have been improved 10 and 8.9 times compared with the initial plasmon detector values respectively. These properties suggest a responsive plasmon detector that can create the same output photocurrent under a lower illuminating power in comparison with conventional plasmon detectors.

## References

1. Hongyu Chen, Hui Liu, Zhiming Zhang, Kai Hu, and Xiaosheng Fang, "Nanostructured Photodetectors: From Ultraviolet to Terahertz," Adv. Mater, vol. 28, pp. 403–433, 2016.
2. P. Berini, "Surface plasmon photodetectors and their applications," Laser Photonics Rev., DOI 10.1002/lpor.201300019, 2013.
3. M. Alavirad, A. Olivieri, L. Roy, P. Berini, "High-responsivity sub-bandgap hot-hole plasmonic Schottky detectors," Opt. Express, vol. 24, pp. 22544-54, 2016.
4. T. Ishi, J. Fujikata, K. Makita, T. Baba, and K. Ohashi, "Si nano-photodiode with a surface plasmon antenna," Jpn. J. Appl. Phys., vol. 44, L364-L366, 2005.
5. J. Rosenburg, R. V. Shenoi, T. E. Vandervelde, S. Krishna, and O. Painter, "A multispectral and polarization-selective surface-plasmon resonant midinfrared detector," Appl. Phys. Lett., vol. 95, 161101, 2009.
6. Akbari, R. N. Tait, and P. Berini, "Surface plasmon waveguide Schottky detector," Opt. Express 18, pp. 8505– 8514, 2010.
7. A. M. Livani and H. Kaatuzian, "Modulation–Frequency Analysis of an Electrically Pumped Plasmonic Amplifier," Plasmonics, vol. 12, Issue 1, pp 27–32, 2017.
8. M. Keshavarz Moazzam, H. Kaatuzian, "Design and Investigation of a N-type Metal/Insulator/Semiconductor/Metal structure 2-port Electro-Plasmonic addressed Routing Switch," Applied Optics, Vol. 54, No. 20, PP. 6199 - 6207, 03 July 2015.
9. C. Scales, I. Breukelaar, R. Charbonneau, P. Berini, "Infrared Performance of Symmetric Surface-Plasmon Waveguide Schottky Detectors in Si," J. Lightwave Technol. 29, pp. 1852–1860, 2011.
10. Goykhman, B. Desiatov, J. Khurgin, J. Shappir, U. Levy, "Waveguide based compact silicon Schottky photodetector with enhanced responsivity in the telecom spectral band," Opt. Express 20, pp. 28594–28602, 2012.
11. S. Muehlbrandt, A. Melikyan, T. Harter, K. Köhnle, A. Muslija, P. Vincze, S. Wolf, P. Jakobs, Y. Fedoryshyn, W. Freude, J. Leuthold, C. Koos, & M. Kohl, "Silicon-plasmonic internal-photoemission detector for 40 Gbit/s data reception," OPTICA Optical Society of America, Vol. 3, No. 7, pp. 741-747, 2016.
12. Yannick Salamin, Ping Ma, Benedikt Baeuerle, Alexandros Emboras, Yuriy Fedoryshyn, Wolfgang Heni, Bojun Cheng, Arne Josten, and Juerg Leuthold, "100 GHz Plasmonic Photodetector," ACS photonics, 5, pp. 3291−3297, 2018.
13. P. Ashburn, SiGe Heterojunction Bipolar Transistors, John Wiley &Sons, 2003.
14. E. Rastegar Pashaki, H. Kaatuzian, A. Mallah Livani, H. Ghodsi, " Design and investigation of a balanced silicon-based plasmonic internal-photoemission detector," Applied Physics B Laser and optics, https://doi.org/10.1007/s00340-018-7111-x, 2018.
15. Silvaco, Inc. (2013, Oct. 2). Atlas User's Manual [online]. Available: http://www.silvaco.com.
16. Lang, D.V., "Measurement of the band gap of $Ge_xSi_{1-X}$/Si strained-layer hetero-structures," Appl. Phys. Lett., Vol. 47, No. 12, pp. 1333-1335, 1985.
17. S.M. Sze, Physics of semiconductor devices, Wiley-Inter science publication, chap. 5, 2nd ed., 1981.
18. Apanovich Y. et. al, "Numerical Simulation of Sub-micrometer Devices, Including Coupled Non-local Transport and Non-isothermal Effects," IEEE Trans. Electron Devices Vol. 42, No. 5, pp. 890-898, 1995.
19. Apanovich Y. et. al, "Steady-State and Transient Analysis of Sub-micron Devices Using Energy Balance and Simplified Hydrodynamic Models," IEEE Trans. Comp. Aided Design of Integrated Circuits and Systems Vol. 13, No. 6, pp. 702-710, June 1994.
20. D. J. Richey, J. D. Cressler and A. J. Joseph, "Scaling issues and profile optimization in advanced UHV/CVD SiGe HBTs," IEEE Trans. Elect. Dev., vol. 44, pp. 431-440, 1997.
21. J. D. Cressler and N. Guofu, Silicon Germanium Heterojunction Bipolar Transistors, Boston-London: Artech House, 2002.

22. Matsuzawa, K., K. Uchida, and A. Nishiyama, "A Unified Simulation of Schottky and Ohmic Contacts", IEEE Trans. Electron Devices, Vol. 47, No. 1, pp. 103-108, Jan. 2000.
23. Ieong, M., Solomon, P., Laux, S., Wong, H., and Chidambarro, D., "Comparison of Raised and Schottky Source/Drain MOSFETs Using a Novel Tunneling Contact Model", Proceedings of IEDM, pp. 733-736, 1998.
24. Christine Scales, and Pierre Berini, "Thin-Film Schottky Barrier Photodetector Models," IEEE Journal of Quantum electronics, vol. 46, no. 5, pp. 633-643, May 2010.
25. Marco Bernardi, Jamal Mustafa, Jeffrey B. Neaton & Steven G. Louie, "Theory and computation of hot carriers generated by surface plasmon polaritons in noble metals," nature communications, DOI: 10.1038/ncomms8044, Jun. 2015.
26. F. M. Bufler, P. Graf, B. Meinerzhagen, D. Adeline, M. M. Rieger and H. Kabbel, "Low- and high-field electron-transport parameters for unstrained and strained Si1-xGex," IEEE Elect. Dev.Lett., vol. 18, pp. 264-266, 1997.
27. H. Kaatuzian, Photonics, AmirKabir university press, Vol. 1, chap. 2, 5'th printing 2017.